\documentclass[preprintnumbers,amssymb,amsmath,superscriptaddress,letterpaper,nofootinbib,twocolumn,floatfix]{revtex4}[10pt]

\usepackage{graphicx}
\usepackage{dcolumn}
\usepackage{bm}
\usepackage{natbib}
\usepackage{calligra}
\usepackage[T1]{fontenc}
\usepackage{egothic}
\usepackage[T1]{fontenc}
\newfont{\rsfsten}{rsfs10 scaled 1200}
\newfont{\rsfsseven}{rsfs10 scaled 1200}
\newfont{\rsfsfive}{rsfs10 scaled 1200}
\usepackage{epsfig}
\usepackage{units}

\newcommand{\be}{\begin{equation}}
\newcommand{\ee}{\end{equation}}
\newcommand{\bea}{\begin{eqnarray}}
\newcommand{\eea}{\end{eqnarray}}

\def\lsim{\mathrel{\raise.3ex\hbox{$<$\kern-.75em\lower1ex\hbox{$\sim$}}}}
\def\gsim{\mathrel{\raise.3ex\hbox{$>$\kern-.75em\lower1ex\hbox{$\sim$}}}}

\begin{document}

\hspace*{130mm}{\large \tt FERMILAB-14-312-A}
\vskip 0.2in


\title{Improving the Sensitivity of Gamma-Ray Telescopes to Dark Matter Annihilation in Dwarf Spheroidal Galaxies}

\author{Eric Carlson}
\affiliation{Department of Physics, University of California, Santa Cruz, CA}

\author{Dan Hooper}
\affiliation{Fermi National Accelerator Laboratory, Center for Particle Astrophysics, Batavia, IL}
\affiliation{University of Chicago, Department of Astronomy and Astrophysics Chicago, IL }

\author{Tim Linden}
\affiliation{Kavli Institute for Cosmological Physics University of Chicago, Chicago, IL }

\begin{abstract}
The Fermi-LAT collaboration has studied the gamma-ray emission from a stacked population of dwarf spheroidal galaxies and used this information to set constraints on the dark matter annihilation cross section. Interestingly, their analysis uncovered an excess with a test statistic (TS) of 8.7. If interpreted naively, this constitutes a 2.95$\sigma$ local excess ($p$-value=0.003), relative to the expectations of their background model. In order to further test this interpretation, the Fermi-LAT team studied a large number of blank sky locations and found TS>8.7 excesses to be more common than predicted by their background model, decreasing the significance of their dwarf excess to 2.2$\sigma$ ($p$-value=0.027). We argue that these TS>8.7 blank sky locations are largely the result of unresolved blazars, radio galaxies, and starforming galaxies, and show that multi-wavelength information can be used to reduce the degree to which such sources contaminate the otherwise blank sky. In particular, we show that masking regions of the sky that lie within $1^{\circ}$ of sources contained in the BZCAT or CRATES catalogs reduces the fraction of blank sky locations with TS>8.7 by more than a factor of two. Taking such multi-wavelength information into account can enable experiments such as Fermi to better characterize their backgrounds and increase their sensitivity to dark matter in dwarf galaxies, the most important of which remain largely uncontaminated by unresolved point sources. We also note that for the range of dark matter masses and annihilation cross sections currently being tested by studies of dwarf spheroidal galaxies, simulations predict that Fermi should be able to detect a significant number of dark matter subhalos. These subhalos constitute a population of sub-threshold gamma-ray point sources and represent an irreducible background for searches for dark matter annihilation in dwarf galaxies.

\end{abstract}

\maketitle


The Milky Way's dwarf spheroidal galaxies (dSphs) represent a very promising set of targets for indirect dark matter searches. Although the flux of gamma-rays from dark matter annihilating in these systems is predicted to be considerably lower than from the Galactic Center, the lower astrophysical backgrounds make dSphs comparably sensitive to annihilating dark matter. Furthermore, precision stellar rotation measurements have been used to directly constrain the dark matter density profiles of many dSphs, making it possible to predict the dark matter annihilation rate within such systems.

Several groups have analyzed dSphs as observed by the Fermi-LAT~\citep{Atwood:2009ez,GeringerSameth:2011iw, Ackermann:2011wa,Ackermann:2012nb,Ackermann:2013yva}. The most recent of these efforts was carried out by the Fermi-LAT collaboration, which investigated a stacked population of 25 dSphs, using four years of data~\citep{Ackermann:2013yva}. The expected sensitivity of this analysis was sufficient to exclude dark matter with an annihilation cross section equal to the standard estimate for a thermal relic ($\sigma v\simeq 2-3 \times 10^{-26}$ cm$^3$/s) for masses below $m_{\rm DM} \sim 90$ GeV in the case of annihilation to $b\bar{b}$. If this expected sensitivity had been realized, the resulting limit would have been the most stringent to date, exceeding those derived from gamma-ray observations of the Galactic Center~\cite{Hooper:2012sr}, galaxy clusters~\cite{2010JCAP...05..025A,2010JCAP...12..015D}, or the isotropic gamma-ray background~\cite{2010JCAP...04..014A,2010JCAP...11..041A,Cholis:2013ena}. However, the actual limit obtained by this analysis was significantly weaker than expected (by a factor of $\sim$4-5 for $m_{\rm DM} \simeq10-100$ GeV). The difference between the expected and actual limits was greatest for a dark matter particle of mass $m_{\rm DM}\sim$~25~GeV annihilating to $b\bar{b}$. At this mass, an excess corresponding to a test statistic (TS) of 8.7 was found. Interestingly, the normalization and spectral shape of this excess are consistent with those produced by dark matter models capable of accounting for the gamma-ray signal observed from the Galactic Center (e.g. with $m_{\rm DM}\sim$~30-40~GeV and a cross section of $\sigma v$~=~(1.7 -- 2.3)~$\times$~10$^{-26}$~cm$^3$s$^{-1}$ to $b\bar{b}$~\cite{Daylan:2014rsa,Goodenough:2009gk,Hooper:2010mq,Abazajian:2012pn,Gordon:2013vta,Hooper:2013rwa,Abazajian:2014fta,Calore:2014xka}).

If one assumes that the astrophysical emission models employed by the Fermi-LAT team are entirely accurate (to the level of Poisson noise), a TS=8.7 excess would correspond to a local significance of 2.95$\sigma$. This level of accuracy, however, is not expected for current astrophysical background models. In order to empirically quantify this mismodeling, the Fermi-LAT team studied 7500 random ``blank sky'' locations at galactic latitudes comparable to the dSph population (|b|~$>$~30$^\circ$) and at least 1$^\circ$ (5$^\circ$) from any point source (extended source) in the 2FGL catalog~\citep{Fermi-LAT:2011iqa}. They then calculated the TS value obtained by placing a mock dSph at each location, and used the probability distribution of these residuals to convert the TS value of the dSph analysis into a significance. This method found ``blank sky'' locations yielding TS>8.7 to be more common (by a factor of 8.9) than predicted by the background model. When this is taken into account, the statistical significance of the measured dSph excess is reduced to a local value of 2.2$\sigma$.\footnote{The Fermi-LAT collaboration analysis includes a trials factor of approximately 3, due to the multiplicity of dark matter models they test. However, if dSphs are being studied in order to confirm or exclude a dark matter interpretation of the signal observed from the Galactic Center, then this trials factor is irrelevant. In this paper we only consider the local significance of the dwarf excess.}

\begin{figure}[!t]
\includegraphics[width=3.4in,angle=0]{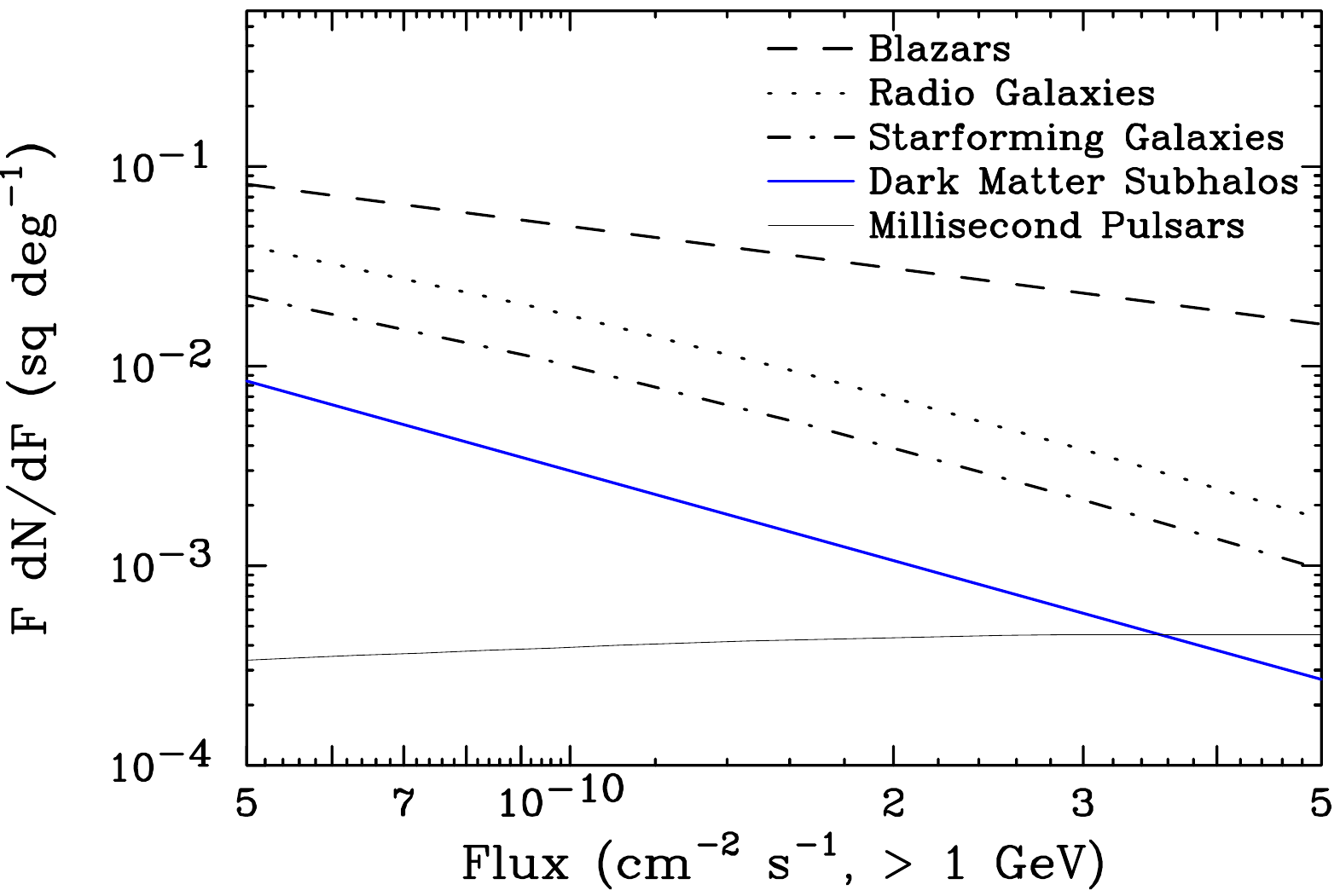}
\caption{A set of empirically driven population models for a number of gamma-ray source classes, including blazars~\cite{Collaboration:2010gqa}, radio galaxies~\cite{Inoue:2011bm}, star forming galaxies~\cite{2012ApJ...755..164A}, and millisecond pulsars~\cite{Hooper:2013nhl}. Also shown is an estimate for the distribution of dark matter subhalos, calculated as in Ref.~\cite{Berlin:2013dva} (using the mass-concentration relationship of Ref.~\cite{Sanchez-Conde:2013yxa}) for the case of $m_{\rm DM}=$35 GeV and $\sigma v=2\times10^{-26}$ cm$^3$/s to $b\bar{b}$.}
\label{fig:TS_source_classes}
\end{figure}

The Fermi Collaboration has been non-committal regarding the departures of their background model from the observed distribution, mentioning both the presence of unresolved point sources and imperfect diffuse background modeling as possible factors~\citep{Ackermann:2013yva}. To estimate the contribution to this deviation from unresolved point sources, we have considered empirically constrained population models for a number of gamma-ray source classes, including blazars~\cite{Collaboration:2010gqa}, radio galaxies~\cite{Inoue:2011bm}, star forming galaxies~\cite{2012ApJ...755..164A}, and millisecond pulsars~\cite{Hooper:2013nhl}. In Fig.~\ref{fig:TS_source_classes} we plot the flux distribution predicted for these source population models, in the range likely to lead to TS~$\sim 8.7$ departures from the background model. Although radio galaxies and starforming galaxies are each predicted to provide non-negligible contributions to Fermi's unresolved source population, blazars constitute the largest number of such sources. This is not surprising given that blazars are the most numerous point sources in the high-latitude gamma-ray sky and are thought to be responsible for the majority of the anisotropy observed in the extragalactic gamma-ray background (which is dominated by sources just below the Fermi-LAT point source detection threshold)~\citep{Ackermann:2012uf, Abazajian:2010pc, Venters:2010bq, Venters:2011gg, Cuoco:2012yf, Harding:2012gk}.

\begin{figure}[!t]
\includegraphics[width=3.4in,angle=0]{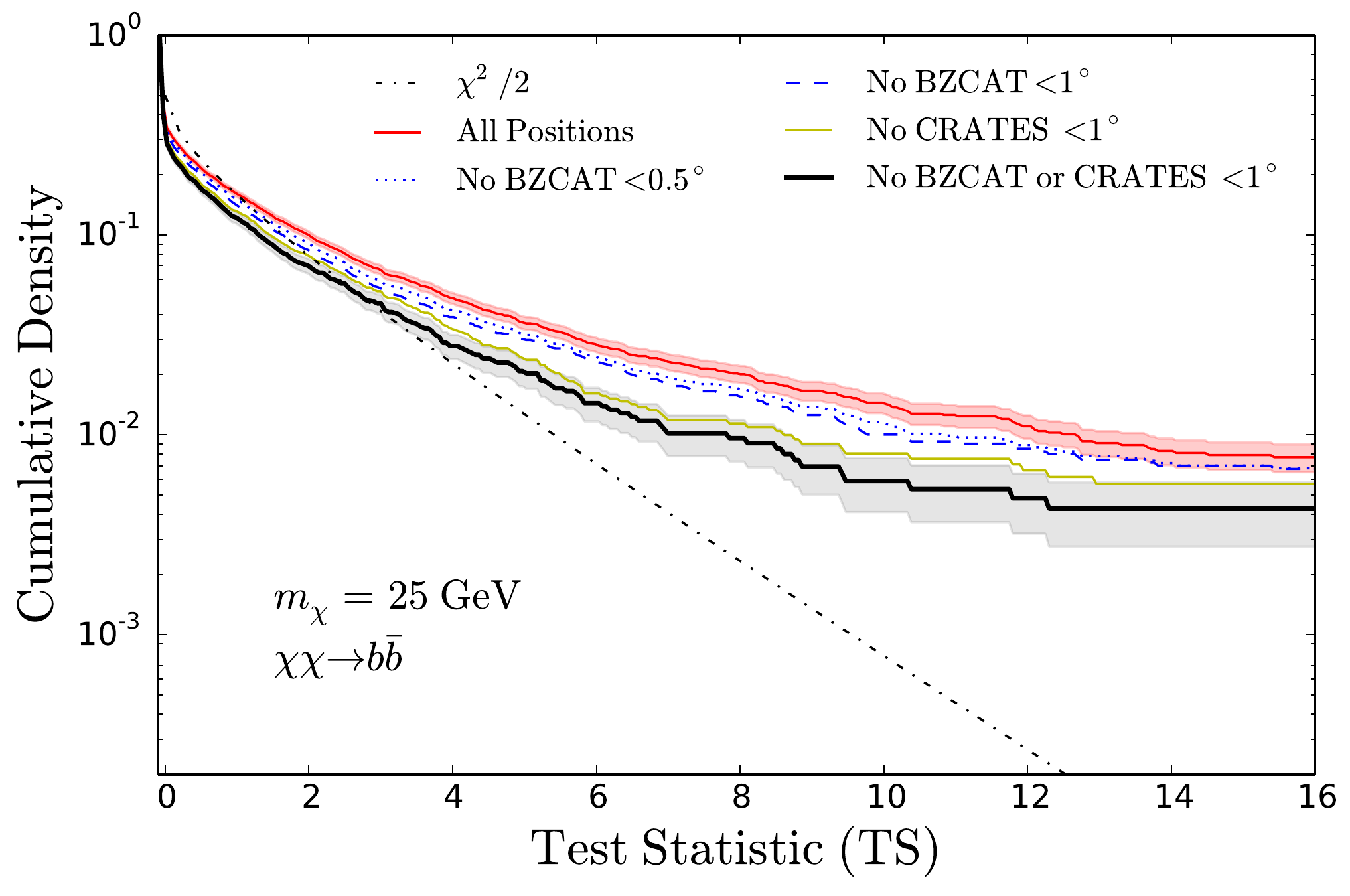}
\caption{The distribution of test statistic (TS) values for a population of 5200 randomly selected sky locations constrained to lie at a galactic latitude |b|~$>$~30$^\circ$ and at least 1$^\circ$ ($5^{\circ}$) away from point-like (extended) 2FGL sources. We show the distribution of TS values when we no additional sky regions are excluded from our analysis (red solid with shaded poisson error bars), and when we make regions that lie within 0.5$^\circ$ of a BZCAT source (blue dotted), within 1$^\circ$ of a BZCAT source (blue dotted), within 1$^\circ$ of a CRATES source (yellow solid), and within 1$^\circ$ of either the BZCAT or CRATES sources (black solid). In the case that sky locations are constrained to lie at least 1$^\circ$ away from any BZCAT or CRATES source, the density of TS~$>$~8.7 locations is reduced by a factor of 2.1.}
\label{fig:TSdist}
\end{figure}

To estimate the impact of these unresolved sources on the Fermi dSph analysis, we simulated the gamma-ray signal from the unresolved source model shown in Fig.~\ref{fig:TS_source_classes} assuming a $dN/dE \propto E^{-2.2}$ spectral shape for blazars, radio galaxies and starforming galaxies. This simulation found that these unresolved sources could account for approximately $\sim$80\% of the TS$>8.7$ ``blank sky'' locations observed in the Fermi dSph analysis. Although uncertainties in the blazar, radio galaxy, and starforming galaxy population models are significant, this calculation leads us to conclude that unresolved sources are likely to be responsible for most of the observed deviations from Fermi's diffuse background model.

\begin{figure*}[!]
\includegraphics[width=7.1in,angle=0]{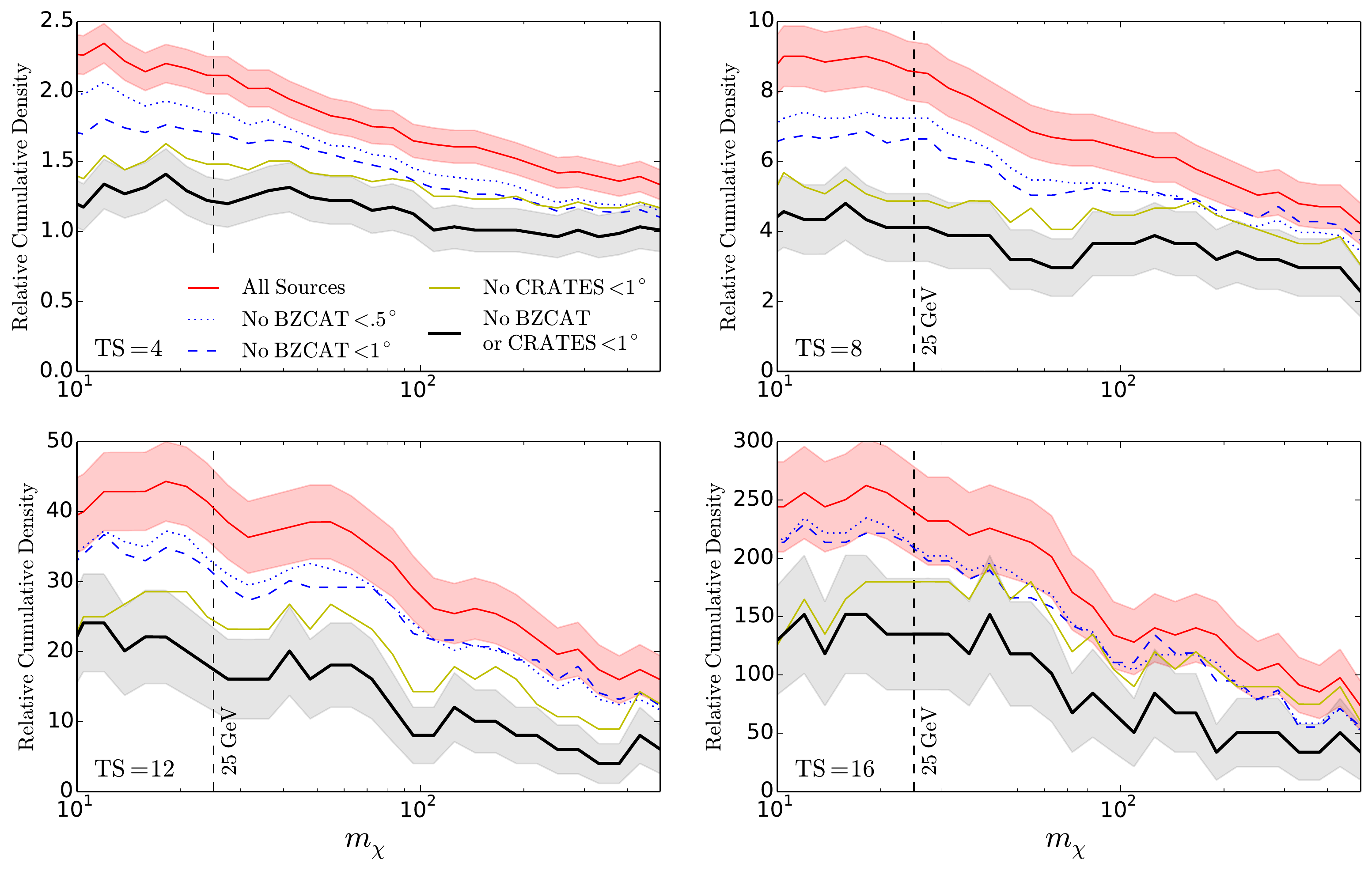}
\caption{The ratio of the fraction of ``blank sky'' locations (at |b|~$>$~30$^\circ$) with TS$>$\{4, 8, 12, 16\} compared to that predicted by Fermi's diffuse emission model, as a function of dark matter mass (assuming annihilations to $b\bar{b}$). For all curves, the sky locations are chosen to be at least 1$^\circ$ (5$^\circ$) from 2FGL point (extended) sources.  This ratio is reduced when we further mask around BZCAT and CRATES sources, as denoted in the key. Shaded regions represent Poisson errors. By masking regions near BZCAT and CRATES sources, we can significantly reduce the fraction of the sky with TS values larger than predicted by Fermi's diffuse model. This conclusion is true for spectral shapes corresponding to wide range of dark matter masses.}
\label{fig:DMmass}
\end{figure*}

We can utilize multi-wavelength information to reduce the impact of unresolved sources on Fermi's dSph analysis. Notably, radio surveys have located a significant population of blazars, star-forming galaxies, radio galaxies, and pulsars which do not appear in the 2FGL catalog, but that are nonetheless likely to be significant gamma-ray emitters. Such sources will appear in the Fermi analysis as small departures from the background model.

In this letter, we utilize two multi-wavelength source catalogs. The first of these is the Roma-BZCAT Multi-Frequency Catalog of Blazars (BZCAT), which currently contains 3149 known blazar sources~\citep{Massaro:2008ye}, 2274 of which are located at high galactic latitude (|b|~$>$~30$^\circ$).\footnote{The BZCAT catalog contains blazars detected by multiple surveys, and has a highly anisotropic sensitivity. For example, the catalog contains 1472 sources with b~$>$~30$^\circ$ and only 802 sources with b~$<$~-30$^\circ$.} Second, we make use of the more than 11,000 bright flat-spectrum radio sources observed by the Combined Radio All-Sky Targeted Eight-GHz Survey (CRATES)~\cite{Healey:2007by}. CRATES claims an all-sky exposure down to 65~mJy at 4.8~GHz. While the nature of these sources is not classified, their spectra are often consistent with source classes likely to produce significant gamma-ray emission.

The strategy we propose here is to use the information provided by BZCAT and CRATES to select regions of the ``blank sky'' that are the least likely to contain significant emission from unresolved gamma-ray point sources. To study the impact of such an approach, we use 4 years of Fermi-LAT data and calculate the distribution of TS values obtained for a set of 5,200 high-latitude (|b|~$>$~30$^\circ$) blank sky locations, each chosen to lie at least \{0$^\circ$, 0.5$^\circ$, 1$^\circ$\} from the nearest BZCAT or CRATES source. For each location, we extract the Fermi-LAT data using photons from the {\tt P7REP$\_$CLEAN} event class, using standard analysis cuts.\footnote{${\tt DATA\_QUAL=1~\& \&~ LAT\_CONFIG=1~\& \&~ ABS(ROCK\_ANGLE)<52}$}
To calculate the TS for each location, we employ the {\tt gtlike} tool utilizing the {\tt MINUIT} algorithm to create a best fit model, including  a mock point source at the chosen location, as well as all 2FGL sources and the {\tt P7v15} and {\tt P7REP\_CLEAN\_V15} diffuse and isotropic background models. We note that in the Fermi-LAT analysis, the mock sources are not point sources, but instead include 300 realizations of each dSph~\citep{Ackermann:2013yva}. However, the Fermi-LAT collaboration notes that this has only a marginal effect on the calculated TS for each source. We have confirmed this result and find that our measurement falls within the statistical errors of the Fermi-LAT measurement when no BZCAT or CRATES sources are masked. In Fig.~\ref{fig:TSdist}, we plot the cumulative distribution of TS values for the different masking choices, assuming that the mock sources have a spectrum equivalent to a 25~GeV dark matter particle annihilating to $b\bar{b}$, calculated using \texttt{PYTHIA 8.183}~\cite{Sjostrand:2007gs}. When considering only ``blank sky'' locations more than $1^{\circ}$ from any BZCAT or CRATES source, the diffuse background model provides a much better description of the data. In particular, for the case shown in Fig.~\ref{fig:TSdist}, the cumulative density of TS>8.7 residuals is reduced by a factor of 2.1 after applying this cut. This effect modestly increases the significance of the TS=8.7 excess observed by the Fermi collaboration from 2.2$\sigma$ to 2.5$\sigma$. In Fig.~\ref{fig:DMmass}, we show the impact of these cuts as a function of the dark matter mass. 

Of course, the correction described in the previous paragraph can only be self-consistently applied to the excess found in Ref.~\citep{Ackermann:2013yva} if we ensure that the dSph fields are not also contaminated by BZCAT or CRATES sources. Notably, the Fermi-LAT team reanalyzed the regions of interest around each dSph, and found no new point sources within 1$^\circ$, decreasing the likelihood that any bright sources are contaminating the dwarf
analysis.

In Table~\ref{tab:BZCAT} we list the dSphs used in the Fermi-LAT analysis which are located within $1^{\circ}$ of at least one BZCAT or CRATES source.  Of most interest are the three dSphs which dominate the excess observed by Fermi:  Segue 1, Usra Major II, and Willman 1~~\citep{Ackermann:2013yva}. Although, these three dSphs each have one BZCAT or CRATES source within this radius, none of these sources are particularly nearby (all are $>0.7^{\circ}$ away). In order to test the impact of these three sources, we utilize the Fermi tools and calculate their TS values to be 0.00, 4.23 and 9.71 for J100955+160223 (0.70$^\circ$ from Segue 1), J0854+6218 (0.91$^\circ$ from Ursa Major II) and J1048+5009 (0.87$^\circ$ from Willman 1), respectively. In order to estimate the TS value of these sources as evaluated at the location of the dSphs under investigation, we produce 50 simulations for a location 0.7$^\circ$ from a simulated TS=10 source. These simulations revealed that the residual TS from the misidentification of the source is TS$\lsim$1, corresponding to $\lsim$10\% of the actual source TS. This indicates that these three BZCAT and CRATES sources are unlikely to be responsible for a significant fraction of the dSph excess. However, we note that a more thorough re-analysis of the dSph population should investigate the potential for low-TS emission from this population.

Finally, the fraction of blank sky locations with higher than expected TS values may also include a contribution from dark matter subhalos. Most important for the case at hand are those subhalos with masses just below those of the dSphs themselves, which are universally predicted by numerical simulations~\citep{Springel:2008cc,Diemand:2008in}. These sources are expected to be distributed nearly isotropically across the sky, with an angular extent that is generally much smaller than the Fermi-LAT point-spread function. In Fig.~\ref{fig:TS_source_classes}, we show the flux distribution of such sources, as calculated in Ref.~\cite{Berlin:2013dva} (but updated using the mass-concentration relationship of Ref.~\cite{Sanchez-Conde:2013yxa}), for the case of $m_{\rm DM}=$35 GeV and $\sigma v=2\times10^{-26}$ cm$^3$/s to $b\bar{b}$. From this figure, we see that while dark matter subhalos are unlikely to dominate Fermi's unresolved source population, they may represent a significant class of unassociated gamma-ray sources. This population is qualitatively different from that of blazars or radio bright galaxies in that while the latter sources constitute a background that could be effectively eliminated using multi-wavelength information, the former corresponds to an irreducible background, with a predicted luminosity that is directly proportional to that of the dSphs being investigated. The ``blank-sky" background modeling employed in Fermi's dSph analysis naturally includes regions of the sky populated by such subhalos, potentially producing an excess of high-TS sources \emph{because} of the existence of a dark matter annihilation signal, rather than in lieu of it.

\begin{table}[t]
\footnotesize
\begin{tabular}{|c|c|}
\hline

dSph   &       Nearby Blazars (Distance to dSph $^\circ$) \\
\hline \hline
 Bootes 1   &   J1359+1436 (.13) \,\, J1401+1350 (.74)  \\
&J140136+151303 (0.80)  \\
\hline
Bootes 3   &   J135948+270834 (0.71)\\
\hline
Canes Venatici 1   &    J132457+325160 (0.97)\\
\hline
Draco &   J1715+5724 (0.85)   \\
\hline
Hercules   &  J162737+121550 (0.95)  \\
\hline
Leo 4  &   J1133+0015 (0.80) \,\,J113631-005250 (0.98) \\
\hline
Leo 5   &   J1131+0234 (.40) \,\,J1132+0237 (.58) \\
&J112940+021817 (0.38)   \\
\hline
Pisces 2  & J225823+051634 (0.68)  \,\, J230153+060906 (0.87)\\
\hline
Sculptor   &   J0100-3337* (0.04) \,\,J010107-334758 (0.24)\\
& J005817-334755 (0.41)\,\, J005819-341957 (0.76) \\
& J010307-342458 (0.97)  \\
\hline
Segue 1   &   J100955+160223 (0.70) \\
\hline
Sextans &   J1010-0200* (0.70) \,\, J101454-005506 (0.82)  \\
\hline
Ursa Major 1   &   J103034+513236 (0.77)\\
\hline
Ursa Major 2   &   J0854+6218* (0.91)  \\
\hline
Willman 1   &   J1048+5009 (0.87)  \\
\hline \hline
\end{tabular}
\caption{A list of BZCAT and CRATES sources that lie within 1$^{\circ}$ of a dwarf spheroidal galaxy studied by Fermi-LAT team~\citep{Ackermann:2013yva}. The distance to the source (in degrees) is given in parentheses. Any source detected in both catalogs is listed with the BZCAT coordinates and marked with an asterisk.}
\label{tab:BZCAT}
\end{table}

In this \emph{letter}, we have investigated three effects that may alter the \emph{interpretation} of the TS=8.7 excess observed in the stacked population of dSphs by the Fermi-LAT Collaboration in Ref.~\citep{Ackermann:2013yva}:

\begin{itemize}
 \item We show that more than 50\% of the TS$>$8.7 residuals observed in blank sky locations by Fermi are the result of sources identified in the BZCAT and CRATES catalogs. Recent population models of blazars, radio galaxies, and starforming galaxies lead us to expect that an even greater fraction of such residuals are the result of unresolved point sources. 
\item Although BZCAT and CRATES sources are found within 1$^\circ$ of 14 of the 25 dSphs analyzed in Ref.~\citep{Ackermann:2013yva}, the three dSphs most responsible for the observed excess (Segue 1, Ursa Major II, Willman 1) have no such sources within $0.7^{\circ}$, making them unlikely to be highly contaminated. 
  \item For the range of dark matter masses and cross sections currently being probed by gamma-ray observations of dSphs, one expects a flux distribution of dark matter subhalos that would account for $\sim$5-10\% of the unresolved source population. Even if all astrophysical sources are accurately modeled, these subhalos will constitute an irreducible background for gamma-ray studies of dSphs. 
\end{itemize}

{\bf Acknowledgements.}
We thank Keith Bechtol and Alex Drlica-Wagner for helpful comments. TL is supported by the National Aeronautics and Space Administration through Einstein Postdoctoral Fellowship Award No. PF3-140110.  EC is supported by a NASA Graduate Research Fellowship under NASA NESSF Grant No. NNX13AO63H.

\bibliographystyle{h-physrev.bst}

\bibliography{DwarfTS}

\end{document}